\documentstyle[aps,amssymb,eqsecnum,epsf]{revtex}

\begin{document}

%\normalsize\textlineskip
%\thispagestyle{empty}
%\setcounter{page}{1}

%\copyrightheading{}			%{Vol. 0, No. 0 (1993) 000--000}

%\vspace*{0.88truein}

%\fpage{1}

\centerline{\bf Matching characteristic codes: exploiting two directions}
\vspace*{0.37truein}
\centerline{Luis Lehner}
\vspace*{0.015truein}
\centerline{\footnotesize\it Center for Relativity, The University
of Texas at Austin}
\baselineskip=10pt
\centerline{\footnotesize\it Austin, TX 78712,
USA.}
%\vspace*{0.225truein}
%\publisher{(received date)}{(revised date)}

\vspace*{0.21truein}
\begin{abstract}
Combining incoming and outgoing characteristic formulations can
provide numerical relativists with a natural implementation of 
Einstein's equations that better exploits the causal properties of the
spacetime and gives access to both null infinity and the
interior region simultaneously (assuming the foliation
is free of caustics and crossovers). We discuss
how this combination can be performed and illustrate its behavior in the 
Einstein-Klein-Gordon field in 1D.
\end{abstract}

%\pacs{04.25Dm, 04.20Ha, 04.30Db}

%\vspace*{1pt}\textlineskip
\section{Introduction}
\vspace*{-0.5pt}
\label{sec:int}
\noindent
In recent years the characteristic formulation of G.R.\cite{bondi,sachs} has proven
to be a valuable tool for numerical relativists\footnote{For a recent
review on its use in numerical relativity see\cite{jeffnew}.}. Its use has enabled
researchers to: study critical phenomena\cite{garf,hama}; obtain the first `unlimited' evolutions of 
generic single black holes\cite{wobble,movlet}; study
accretion of matter by  black holes\cite{philip,matter} (via a formulation 
based on foliating the spacetime with incoming null hypersurfaces)
and obtain practical access to future null infinity where the gravitational
radiation and other physically relevant quantities are obtained
unambiguously\cite{hpgn,dinve} (via an outgoing
null hypersurfaces foliation).
These investigations adopted either an outgoing or an incoming null hypersurfaces
foliation of the spacetime. Regrettably, in choosing one or the other one has access
to future null infinity or to the black hole
interior (as the incoming null surfaces can penetrate it but not the outgoing!)
but not both. This, 
could be regarded as a drawback for the formulation
as it is desirable in many cases to access both regions; for instance, to describe a
black hole accreting mass from its surrounding and the gravitational waves produced in
the system.

At present, obtaining a model capable of addressing both problems at the same time
requires using Cauchy-characteristic matching (CcM)\cite{dinve,match,manual} or the conformal
formulation of Einstein's equations\cite{helmut,hubner,frauendien}. Unfortunately,
despite much progress in both approaches neither of them is fully working in 3D. 
The main problem facing the CcM approach is the involved set of interpolations
back and forth from a $3D$ cartesian grid to $2$-spheres at the matching worldtube. On the other
hand, the conformal approach has as a drawback its large number of variables which
imposes further constraints on present computer capabilities.

The characteristic formulation\cite{bondi,sachs}, on the other hand, employs a minimum set of
variables ($6$ compared with $12$ in the ADM equations or $53$ in the conformal equations).
The success obtained with it in both formulations suggests that a combination of both could
well be a natural way to study both regions. This approach, which we call
``characteristic-characteristic matching'' ($c^2M$) divides the spacetime in two regions.
The inner region (inside some worldtube $\Gamma$) to be studied  by an incoming
characteristic formulation and the outer region where and outgoing formulation will be used.
Interpolations are carried out at $\Gamma$, however in $3D$ it only involves two-dimensional
interpolations. In a related approach, the {\it double null formulation}\cite{sachsdoubnu},
Einstein equations are also integrated along incoming and outgoing null hypersurfaces and
its implementation might display similar stability properties as well. Unfortunately, to date,
there are no numerical implementations of this approach in 3D but studies on how to carry out
such a task are under development\cite{haywa}. 
Naturally, neither of these approaches is expected to be applicable to all
situations since, for instance, the presence of caustics prevents one to use them. However, in
a caustic-free foliation\footnote{Aside from the natural caustic present at the origin of the
null cones used to coordinatize the incoming null surfaces.},  $c^2M$ represents an enticing proposal
in light of the successful application of the characteristic formulation both in the vacuum
and matter cases. 

We briefly describe the characteristic formulation (both in the incoming
and outgoing formulation) in section 2. Section
3 outlines the matching strategy to be used in this work. We describe
the numerical implementation in section 4 and test it
in section 5. Finally we conclude with some comments
and future research directions.

\section{The characteristic formulation of G.R.}
\label{sec:theory}
\noindent
Einstein's equations can be expressed in notational form as
$G_{ab}=8\pi T_{ab}$. Where $G_{ab}$ is the Einstein tensor and
$T_{ab}$ the stress energy tensor of the matter distribution. In the particular case of
a massless scalar field $\Phi$ coupled to G.R., $T_{ab}$ results\cite{christod}
\begin{equation}
T_{ab} = \nabla_a \Phi \nabla_b \Phi - 1/2 \, g_{ab} \nabla_c \Phi \nabla^c \Phi \, .
\end{equation}

We introduce a coordinate system adapted to either incoming or outgoing 
null hypersurfaces in the following way: the outgoing (incoming) lightlike hypersurfaces
are labeled with
a parameter $u$ ($v$); each null ray on a specific hypersurface
is labeled with $x^A$ $(A=2,3)$ and $r$ is introduced as a surface area
coordinate (i.e. surfaces at $r=const$ have area $4
\pi r^2$) (see Fig. 1). In the resulting $x^a=(u (v),r,x^A)$ coordinates, the metric takes the
Bondi-Sachs form\cite{bondi,sachs}
\begin{eqnarray}
   ds^2 & = & -\left(e^{2\beta}V/r -r^2h_{AB}U^AU^B\right)du^2
        -2e^{2\beta}dudr \nonumber \\ & & -2r^2 h_{AB}U^Bdudx^A 
         +  r^2h_{AB}dx^Adx^B,    \label{eq:bmet}
\end{eqnarray}
for the outgoing case, while for the incoming case:
\begin{eqnarray}
   ds^2 & = & \left(e^{2 \beta} V/  r - r^2 h_{AB}U^AU^B\right)dv^2
        +2e^{2\beta}dudr \nonumber \\ & & -2r^2 h_{AB}U^Bdudx^A 
         +  r^2h_{AB}dx^Adx^B,    \label{eq:bmet_i}
\end{eqnarray}
where in both line elements, $h^{AB}h_{BC}=\delta^A_C$ and $det(h_{AB})=det(q_{AB})$, with
$q_{AB}$ a unit sphere metric. 
Note that the incoming line element, as discussed in\cite{marsa}, can be obtained from the outgoing
version by the substitution $\beta \rightarrow \beta + i \pi/2$. This fact was used in\cite{wobble},
to obtain a $3D$ implementation of the incoming null formulation from the one constructed
for the outgoing case\cite{hpgn}.

\begin{figure}
\epsfxsize=10cm
\centerline{\epsffile{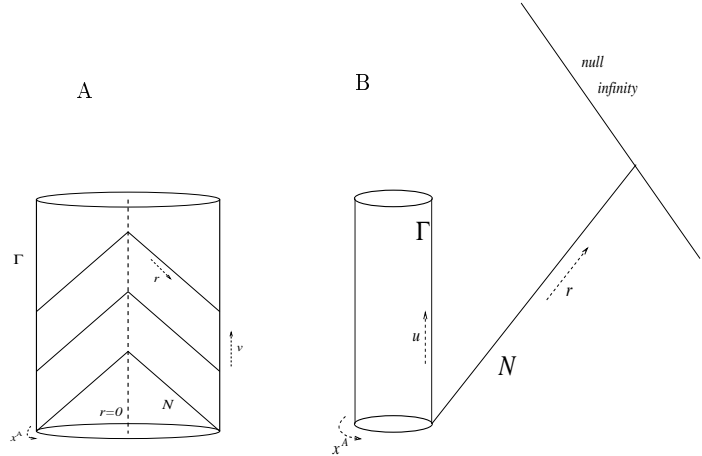}}
\caption{
Incoming (A) and outgoing (B) formulations. In (A), the interior of $\Gamma$ is covered by a
sequence of {\it incoming} light cones. In (B), the exterior of $\Gamma$ is covered by a 
sequence of {\it outgoing} light cones.}
\label{fig:ing}
\end{figure}

In the spherically symmetric case, Einstein's equations and the evolution
equation for the scalar field reduce to:
\begin{eqnarray}
\beta_{,r} &=& 2 \pi \, r \, \Phi_{,r}^2 \, , \label{bet_out} \\
V_{,r} &=& -\, e^{2 \beta} \label{V_out} \, , \\
2 (r \Phi)_{,ru} &=& \frac{1}{r} (r V \Phi_{,r})_{,r}  \, ;\label{phi_out}
\end{eqnarray}
for the outgoing case, while for the incoming one,
\begin{eqnarray}
\tilde \beta_{,r} &=& 2 \pi \, r \, \tilde \Phi_{,r}^2 \, , \label{bet_ingt} \\
\tilde V_{,r} &=& e^{2 \tilde \beta} \label{V_ing} \, , \\
2 (r \tilde \Phi)_{,ru} &=& \frac{1}{r} (r \tilde V \tilde \Phi_{,r})_{,r}  \label{phi_ing}
\end{eqnarray}
(where we distinguish with a tilde the incoming fields from the outgoing ones).

Given that initial data for $\Phi$ ($\tilde \Phi$) on an initial
outgoing (incoming) hypersurface and at some worldtube $\Gamma$ consistent values of
$V$, $\beta$ and $\Phi$ are provided, the equations can be
straightforwardly integrated to yield a unique solution\cite{christod,simoluis}. Consistent
boundary data on $\Gamma$ must satisfy
constraint equations obtained from $G^r_a=8 \pi T^r_a$ ($a=u, x^A$). These
conditions are automatically satisfied when matching across $\Gamma$ to an
interior (exterior) solution generated by the incoming (outgoing) evolution\cite{hpgn}.

Asymptotic quantities of special interest are the Bondi mass and the scalar
news function\cite{scalpower} 
\begin{eqnarray}
M(u) &=& \frac{1}{2} e^{-2 H} r^2 (V/r)_{,r}|_{r=\infty} \label{bondimass} \, , \\
N(u) &=& e^{-2 H} r \Phi_{,u}|_{r=\infty} \label{news} \, ;
\end{eqnarray}
where $H=\beta|_{r=\infty}$. The Bondi mass loss equation is
\begin{equation}
 e^{-2 H} M_{,u} = - 4 \pi N^2
\end{equation}
This last relation provides an important test to the accuracy of our evolution
as
\begin{equation}
M(u=u_o) = M(u=u_f) + 4 \pi \int^{u_f}_{u_o} e^{2 H} N^2  du \, , \label{conservation}
\end{equation}
which we monitor to assess the validity of our numerical implementation.

\section{Matching}
\noindent
Our aim is to match both formulations at a worldtube $\Gamma$ defined
by $r=R=const$ (for simplicity, in fact this condition can be relaxed to $r=R(u)$ and in the $3D$ case
to $R(u,x^A)$). An incoming formulation to be used in the region 
$C_i=\{ (v,r), r \in [0,R_1] \} $ and an outgoing one in the region
$C_o = \{ (u,r), r \in [R_2,\infty) \}$ ($R_2<R<R_1$). By coupling these codes, 
consistent boundary values for
the outgoing case can be obtained from the incoming variable via a
coordinate transformation and viceversa.
In the spherical case, obtaining this coordinate transformation is
straightforward\footnote{The matching strategy in the $3D$ case is naturally more involved and we comment how it
can be implemented in the appendix.}.
 Recall that the incoming line element
is given by
\begin{eqnarray}
ds^2 = e^{2 \tilde \beta} \frac{\tilde V}{\tilde r} dv^2 + 2 e^{2 \tilde \beta} dv d \tilde r
+ \tilde r^2 d \Omega
\end{eqnarray}
Clearly, the transformation
\begin{eqnarray}
dr &=& d \tilde r \\
du &=& dv - \tilde V/ \tilde r \, d \tilde r
\end{eqnarray}
takes the incoming formulation into the outgoing one. This transformation
is valid everywhere and can be used to obtain boundary values by simply 
using the transformation law for a 2-tensor field obtaining:
\begin{eqnarray}
e^{\tilde \beta} = e^{\beta} \, , \, \tilde V = -V \label{transf} \, .
\end{eqnarray}
Additionally, the scalar field must be continuous across $\Gamma$, so the condition
$ \tilde \Phi_\Gamma = \Phi_\Gamma $ applies. Also, in order to ensure
``sheet sources'' produced by discontinuities in derivatives across $\Gamma$
will be not be present one might further impose
\begin{equation}
(k^a \nabla_a \tilde \Phi)^- = (k^a \nabla_a \Phi)^+ \, ;
\end{equation}
where the superscripts $-$ and $+$ denote the regions to the left and right of $\Gamma$ 
respectively and $k^a$ is an arbitrary vector at the worldtube boundary with
non zero radial component.

\subsection{Singularity Excision}
\noindent
In order to avoid dealing with singularities, we use excision to
remove it from our computational domain. This technique first suggested
by Unruh\cite{unruh} is at present the most successful approach to handle singularities.
As it is customary, we excise from the computational domain the region inside
an inner trapping boundary defined by the following.  Given a slice $S$
 on an incoming
null hypersurface ${\cal N}$ described in null coordinates
by $r=R(v,x^A)$, the divergence of the outgoing null normals is given by
\begin{eqnarray}
  {r^2 e^{2\beta}\over 2}\Theta_l &=& -V
           -{1\over \sqrt{q}}[\sqrt{q}(e^{2\beta}h^{AB}R_{,B}-r^2 U^A)]_{,A}
               \nonumber \\ &  & -r(r^{-1}e^{2\beta}h^{AB})_{,r}R_{,A}R_{,B}
                +r^2 U^A_{,r}R_{,A} \, .
    \label{eq:diverg}
\end{eqnarray} 
Hence, the slice is marginally trapped if $\Theta_l=0$.  In the spherically
symmetric case, this reduces to $r=R(v)$, and the divergence of the outgoing null
rays results
\begin{equation}
   F \equiv -2 \, V \, \frac{e^{-2\beta}}{r^2},
    \label{eq:qmarg} 
\end{equation}
Therefore, in our implementation we simply look for the largest $r=r_e$ for which
$F \leq 0$; for $r<r_e$ the field variables are not integrated and simply
set to the values they have at $r_e$. This straightforward procedure excises
the singularity from the computational domain.

\section{Numerical Implementation}
\noindent
\label{sec:numerics}
We introduce a compactified radial
coordinate $x=r/(1+r)$ (so that $x=1 \rightarrow r=\infty$).
Defining the $x$ location of the worldtube by $x_{wt}=R/(1+R)$, we introduce
the outgoing radial grid as
$xo_i=x_{wt} + (i-2) \Delta xo$ where  $\Delta xo = (1-x_{wt})/(N_{xo}-2)$.
Analogously, we define the incoming radial grid as
$xi_j = (j-1) \Delta x_i$ where $\Delta xi = x_{wt}/(N_{xi}-2)$. Notice that
$xo_{2}=xi_{N_{xi-1}}=x_{wt}$. Hence, there is an overlapping region which
will be used to interpolate
from one formulation to the other in order to define boundary values.
Additionally, we choose $v=u$ at $\Gamma$.

\subsection{Evolution Scheme}
\begin{figure}
\centerline{\epsfxsize=150pt\epsfbox{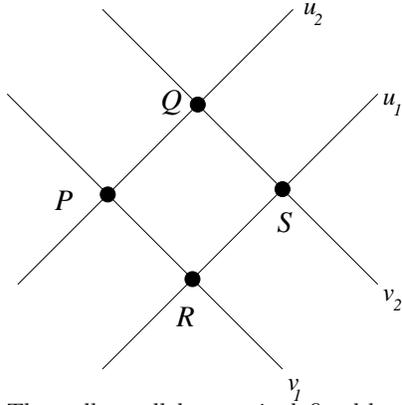}}
\caption{
Parallelogram algorithm set up. The null parallelogram is defined
by the intersection of $2$ outgoing geodesics with $2$ incoming ones.
}
\label{fig:last}
\end{figure}

\noindent
Integration of the evolution equations is done via the parallelogram approach
introduced in\cite{giw}. We here describe  briefly what this method entails for the
outgoing case (it is straightforward to translate it to the incoming case so we do not
expand on it).  Basically, one recognizes that equation (\ref{phi_out}) is equivalent to
\begin{equation}
2 (r \Phi)_{,ur} -\left( \frac{V}{r} (r \Phi)_{,r} \right)_{,r}
   = -\left( \frac{V}{r} \right)_{,r} \Phi \, ;
\end{equation}
defining $G= r \Phi$, the previous equation  can be rewritten as
\begin{equation}
\square G = -\left( \frac{V}{r} \right)_{,r} \frac{G}{r} \, .
\end{equation}
This equation corresponds to the $2$-dimensional wave equation in the
$(u,r)$ plane and can be easily integrated over a null parallelogram
defined by two sets of outgoing and incoming null geodesics (see figure \ref{fig:last}).
Hence, the integral identity 
\begin{equation}
G_Q = G_P +G_R -G_R + \frac{1}{2} \int_A{du dr (\frac{V}{r})_{,r} } \frac{G}{r} \, ,
\end{equation}
can be used to integrate the scalar field up to global second order
accuracy as described in\cite{giw}. Additionally,
the ``hypersurface'' equations are discretized by centered second order differences
yielding a global second order accurate numerical implementation.

\subsection{Matching scheme}
\begin{figure}
\centerline{\epsfysize=170pt\epsfxsize=220pt\epsfbox{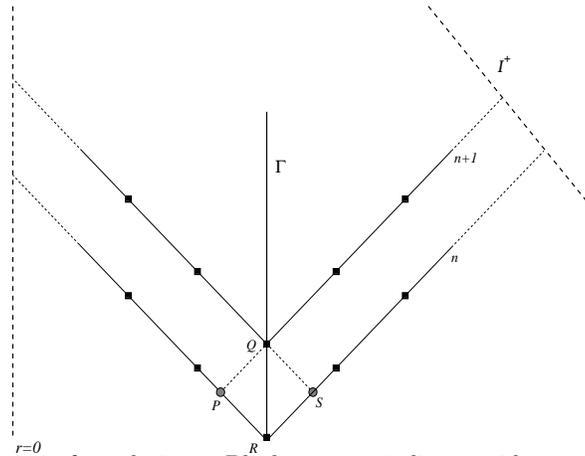}}
\caption{
Matching the characteristic formulations. Black squares indicate grid points while
the grey circles indicate the intersection of incoming and outgoing null rays.
Points $P$, $Q$, $R$ and $S$ define the corners of the parallelogram that will
be used for the integration to obtain the value of $\Phi$ at the $n+1$ level on 
$\Gamma$. 
}
\label{fig:wobble}
\end{figure}

\noindent
A straightforward matching implementation can be obtained
by the following scheme.
Suppose that at time $v=u=u_o (=n \Delta u) $ all field variables are known (refer to figure 3).
In order to obtain starting values at the level $v=u=u_o +\Delta u$ one can find
the intersection of the incoming null surface at $v=u_o+\Delta u$ with the
outgoing null surface corresponding to $u=u_o$, this intersection
can be easily obtained in the following way. First, consider
the outgoing null ray emanating from $x_o = (u_o, x_{wt})$ (R in figure 3);
the intersection point $x_i$ of this ray with the incoming one going
through at $u=u_o$ (indicated by $S$ in figure 3) is obtained by requiring that
\begin{equation}
\int^{x_i}_{x_o} ds^2=0
\end{equation}
which implies that $dr_o =  \frac{V}{2 r} du$. Hence, the values of the
fields at $x_i$ can be easily obtained by interpolation of the field on the outgoing
patch. This value will be used
as starting values to carry out the incoming radial integration. By proceeding
in an analogous way one can get values at $x_o^a = (u_o+\Delta u, R - dr_o)$ ($P$ in figure \ref{fig:wobble})
which provide
starting values for the radial integration on the outgoing null surface.
Additionally, starting values for the hypersurfaces equations are also obtained
by interpolations and applying transformation (\ref{transf}).

To ensure continuity of the field and its radial derivatives we proceed as follows.
Since the intersection of the null surfaces naturally forms a
 null parallelogram (see figure 3), one can
integrate the evolution equation for $\Phi$ (as in section 4.1) and obtain the
value of $\Phi^{n+1}_2$ (since the values
of $\Phi$ at all the other corners are either known directly or can be obtained by interpolation
from the values of the fields at the previous level). This procedure naturally ensures continuity of the field
and its derivative across $\Gamma$.

\section{Tests}
\noindent
Our purpose is to present a model demonstrating the feasibility of a
stable algorithm based upon two regions that cover the spacetime, where
equations are expressed with respect to and incoming or outgoing 
characteristic formulation.
The inner region is evolved using an incoming null algorithm whose
inner boundary lies at the apparent horizon (in case it exists,
otherwise at $r=0$) and whose outer boundary $R_1$ lies beyond the
inner boundary of the outer region (located at $R_2$)
handled by an outgoing null evolution.

\subsection{Scalar wave on a flat background}
\begin{figure}
\centerline{\epsfysize=160pt\epsfxsize=220pt\epsfbox{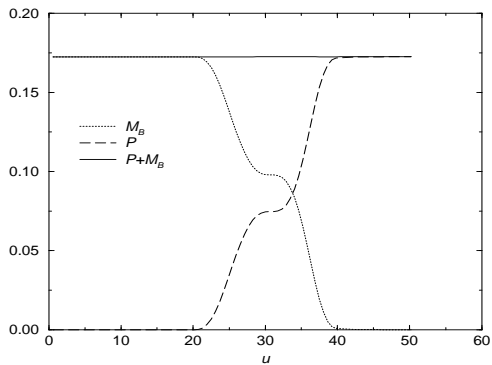}}
\caption{
Conservation of Energy. The plot shows the value of the Bondi mass,
the total radiated energy and the addition of the Bondi mass at a given
time with the energy radiated until that time (indicated by the solid line).
Clearly, the total energy is conserved throughout the evolution. 
}
\label{encons}
\end{figure}

\begin{figure}
\centerline{\epsfysize=200pt\epsfxsize=210pt\epsfbox{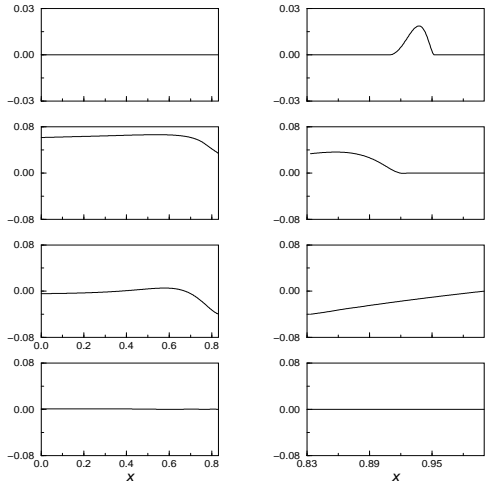}}
\caption{
A sequence of snapshots of the field on both the interior (left) and
exterior region (right). From top to bottom the snapshots correspond to
$u=0,17,34$ and $51$. Initially the scalar field is non zero only
on the exterior region, at later times the field crosses $\Gamma$, reaches
the origin and it is radiated away to future null infinity leaving
behind a flat spacetime.
}
\label{sequence}
\end{figure}

\noindent
As a first test, we choose initial data corresponding to a scalar wave
on a flat background. The initial data for the scalar wave is described
by 
\begin{eqnarray}
\tilde \beta &=& 0 \, ; \, \tilde V = - r \, ; \, \tilde \Phi = 0 \, ;
\end{eqnarray}
%\begin{equation}
%\Phi = \cases { \alpha (1 - r/R_a)^4 (1 - r/R_b)^4 & for r \in [R_a,R_b]   
% & 0 &  otherwise} \label{compact}
%\end{equation}
\begin{equation}
\Phi = \left\{ \begin{array}{ll}
                   \alpha (1 - r/R_a)^4 (1 - r/R_b)^4 & \mbox{for $r \in [R_a,R_b]$ } \\
	            0 &\mbox{otherwise}
	          \end{array}
	  \right.    \label{compact}
\end{equation}

where the values of $\beta$ and $V$ are determined by matching at $R=5$ and integrating
the hypersurface equations. The value of the field at the origin is chosen
so that $\tilde \Phi_{,r} = 0$.
The parameter $\alpha$ was set to be small enough so that the wave would
reach the origin and be dispersed away without a black hole being
formed\footnote{In principle, we could use our numerical implementation to try
to look for critical phenomena as first observed by Choptuik\cite{mattcrit}, we defer such a
study to a future work.}. In our runs, we choose $R_a=10, R_b=20$ and 
study the system for different values of $\alpha$ (with $N_{xi}=N_{xo}=162$). As a test
of our numerical implementation we checked that
equation (\ref{conservation}) was satisfied throughout the evolution.
Our results for the case $\alpha=1$,
are illustrated in figures \ref{encons} and \ref{sequence}.
Energy conservation is satisfied and the evolution of the scalar
field proceeds as expected. The pulse is originally registered at the
outgoing patch; as it travels towards the
origin it crosses $\Gamma$, reaches the origin and it disperses
to infinity radiating all the energy contained in the initial data. 
The spacetime at late times is flat and all the energy of the initial 
pulse has been radiated away.

\subsection{Collapse of a spherically symmetric scalar wave onto a black hole}
\noindent
We choose initial data corresponding to a Schwarzschild black hole
of mass M which is well separated from a localized pulse of 
(mostly) incoming scalar radiation. This problem has been thoroughly
studied in the past by Marsa and Choptuik\cite{marcho} in the Cauchy formulation of G.R. Here,
we do not try to extend their results, rather, to verify the applicability
of $c^2M$.
Our configuration is such that, initially, there is no
scalar field present on the incoming null patch, so the initial
data are simply,
\begin{eqnarray}
\tilde \Phi &=& 0 \, , \, \tilde \beta = 0 \\
\tilde V &=& 2 M - \tilde r .
\end{eqnarray}
On the outgoing patch we choose $\Phi$ as a pulse with
compact support given by (\ref{compact}). 
The values of
$\beta$ and $V$ are determined by matching at $R$ and integrating
the hypersurface equations. Inner boundary conditions
are not needed since the region inside the marginally trapped surface
is excised from the computational domain.

\begin{figure}
\centerline{\epsfysize=160pt\epsfxsize=220pt\epsfbox{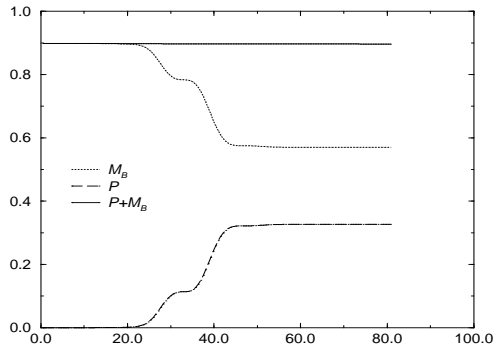}}
\caption{
Conservation of Energy II. The plot shows the value of the Bondi mass,
the total radiated energy and the addition of the Bondi mass at a given
time with the energy radiated until that time (indicated by the solid line).
Values for this run were $\alpha=2$, $R_a = 40M$, $R_b=80M$ and $R=20$.
Part of the initial energy contained in the pulse is radiated away while
part of it is accreted by the black hole. The final spacetime corresponds to a Schwarzschild
spacetime with total mass larger than the initial hole.
}
\label{encons_hole}
\end{figure}

\begin{figure}
\centerline{\epsfysize=180pt\epsfxsize=220pt\epsfbox{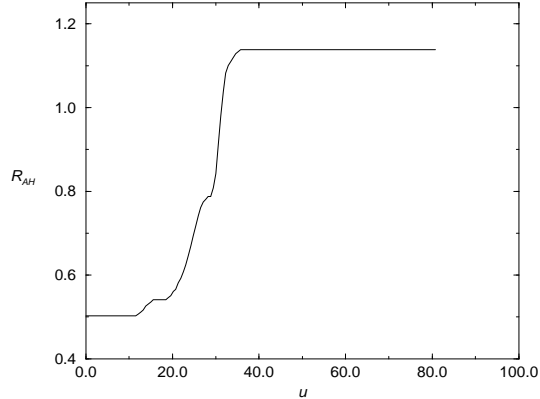}}
\caption{
Apparent horizon location vs. time. The figure shows how the apparent
horizon location grows as energy is accreted by the hole. The final
asymptotic value of $1.1381$ agrees well with twice the value of
the Bondi mass at late times ($2 M_B =1.1388$).
}
\label{mask_hole}
\end{figure}

In our tests, we choose $R=20M$, $R_a=40M, R_b=80M$
and again study the evolution of the system for different values of $\alpha$ (with
$N_{xi}=N_{xo}=162$). As an illustration, we present the results
obtained with $\alpha=2$. As shown in figures \ref{encons_hole} and 
\ref{mask_hole}, the scalar field propagates towards the origin, part
of its ``original'' energy falls
into the hole which increases its mass and the rest is radiated away. 

\begin{figure}
\centerline{\epsfysize=180pt\epsfxsize=180pt\epsfbox{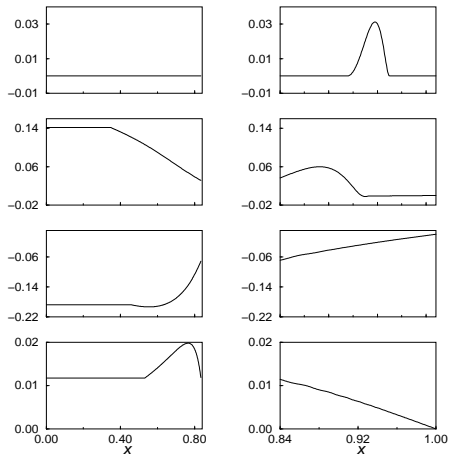}}
\caption{
A sequence of snapshots of the field on both the interior (left) and
exterior region (right). From top to bottom the snapshots correspond to
$u=0,15,30$ and $45$. Initially the scalar field is non zero only
on the exterior region, at later times the field crosses $\Gamma$, and
reached the black hole which accretes some of the energy and so the masked
region increases (shown by the constant lines in the second to fourth plots
on the left). The rest is radiated away to future null infinity leaving
behind a Schwarzschild spacetime with a larger mass. 
}
\label{seq_hole}
\end{figure}

The code successfully 
excises the hole and the area of the apparent horizon reaches an
asymptotic value which agrees with twice the value of the Bondi mass
at late times (see figure \ref{mask_hole}). This result fully agrees with
what is expected, since at late times the spacetime must be describable by a Schwarzschild
metric as a consequence of the no-hair theorems. A sequence of this evolution
is illustrated in figure \ref{seq_hole}. 

\begin{figure}
\centerline{\epsfysize=160pt\epsfxsize=200pt\epsfbox{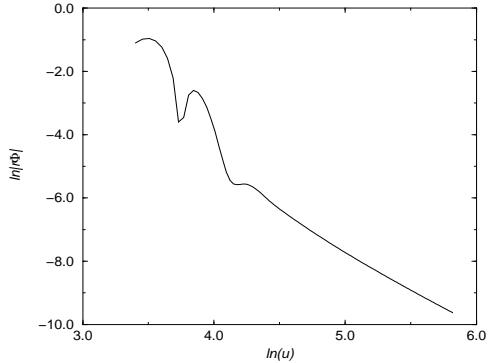}}
\caption{
Late-time power-law tail at null infinity. The measured slope of
$-2.1$ compares well with the theoretical value of $-2$.
}
\label{tails}
\end{figure}

Another important test of the algorithm focuses on general aspects
of wave propagation in a curved background. The qualitative features
of the evolution are well known\cite{waveprop1,waveprop2}. In particular, a
distinctive aspect of wave propagation
on curve backgrounds is that Huygens principle does not apply. In a non-flat
spacetime, radiation backscatters giving rise to quasinormal modes and
tails\cite{price}. Notably, for a given Schwarzschild potential, the power of
the late-time tail is 
characteristic of generic compact support initial data and depends on the 
multipole structure of it. In particular, at future null
infinity the dependence on time goes as $u^{-2-l}$ (with l the multipole index).
In our case, $l=0$ since the initial data corresponds to a spherically symmetric pulse;
the value measured from the simulation is $-2.1$ which agrees with the
expected value of $-2$ (see figure \ref{tails}). The results obtained here reproduce
and, in a way, expand those obtained with Cauchy-characteristic matching\cite{papalaguna}, as
in this case we dynamically find the horizon and use it to excise the hole.
The behavior of the apparent horizon location gives a good indication of
the energy absorbed by the black hole and its area at late times agrees
with that obtained by the asymptotic value of the Bondi mass.

\section{Present and future outlook}
\noindent
This work shows that the matching approach of two characteristic
formulations provides a way of treating both the inner region
and the outer region. At the inner region, it tackles the excision
problem as well as other treatments like Cauchy characteristic
matching and purely Cauchy or characteristic treatments. It has some
advantages over the Cauchy approach because its implementation is much simpler
and computationally cheaper. Its disadvantage though, is that it can
not be applied in the case where caustics might be present in the
integration region. 
Additionally, it provides a means to reach future null infinity, something
that only can be done with the help of a characteristic formulation or
the conformal formulation of Einstein's equations.

This study is the first in a line of research in numerical
relativity aiming to fully match characteristic codes to study matter
coupled to G.R. Preliminary works have shown that, where applicable,
the use of a characteristic evolution can greatly help in investigating this
problem. To date however, these studies have been restricted to either
the incoming or outgoing formulation, thus not being able
to study the inner region or obtaining physical quantities at future
null infinity simultaneously. A combination of both approaches appears to be an ideal
candidate since it would combine the main advantages of both
directions giving access both to future null infinity and to the horizon
region in a natural way. Additionally, it might help ``avoid'' caustics
that can be encountered using the single mapping of each formulation. Just
as several coordinates patches are necessary to describe nontrivial
topologies, a combination of outgoing and incoming coordinates can help to
obtain a smooth mapping of the manifold under study.
The present work represents the first step towards achieving such
a combination. At present, $3D$ 
robust implementations exist for both the incoming and outgoing characteristic
formulations\cite{hpgn,wobble}; further, matter treatments studied
with these formulations have been successfully implemented\cite{philip,matter}.
Thus, demonstrating the feasibility of combining both approaches in $1D$
inspired this work. Future studies will address the matter case
in $1D$ (ie. implementing the fluid equations rather than that of a massless
scalar field) and proceed with full matching in $3D$.

\section{Acknowledgements}
\noindent
This work has been supported by NSF PHY 9800722 and NSF PHY 9800725 to the University of Texas at
Austin. The author wishes to thank Pablo Laguna, Jeffrey Winicour,
Philipos Papadopoulos and Matthew Choptuik for
helpful discussions, Ethan Honda for a careful reading of the manuscript and the National
University of Cordoba for its hospitality where part of this work was completed. 

\section{Appendix}

\noindent
A matching strategy for the general $3D$ case can be devised in a way similar
to the one implemented in the Cauchy-characteristic matching approach.
In the $c^2M$ case, however, its application is simpler since no interpolation
from a $3D$ cartesian grid to $2D$ spheres (slices of the matching worldtube)
is required. This matching technique\cite{manual} is straightforward. First assume
that all field variables are known at hypersurfaces ${\cal N}_{v}$ and ${\cal N}_{u-\Delta u}$
(and earlier ones); that the matching worldtube is located at $\tilde r = R =const$ and
that $u=v$ at $\Gamma$.
The first step which is referred to as {\it extraction} involves obtaining
start-up values (field values at the worldtube) for the outgoing variables at ${\cal N}_u$.
These values are obtained by transforming the incoming metric tensor at
${\cal N}_{v}$ on $\Gamma$ to the outgoing coordinate
system. Instead of solving the geodesic equation to obtain 
this coordinate system one can proceed as follows. Choose an ``affine'' coordinate
system $x^a = (u, \lambda, x^a)$ by 
\begin{equation}
x^a = x^a|_{\Gamma} + \lambda l^a + O(\lambda^2) \,  \; \label{ext_match}
\end{equation}
where $x^a|_{\Gamma} = (v,r=R,\tilde x^a)$; $l^a$ is an outgoing null vector
(normalized by $l^a dv_a = -1$) and we assume the worldtube is at $\tilde r = R = const$.
By inspection, equation (\ref{ext_match}) is the solution to the geodesic
equation with affine parameter $\lambda$ up to second order in $\lambda$. Using this
transformation, one can obtain the metric in outgoing null affine coordinates by
a simple tensor transformation. The final step requires obtaining $r$, as its value
can only be known after the angular components of the outgoing null metric
are obtained, (refer to\cite{manual} for further details).
With these values, one can integrate Einstein's equation and completely determine
the field values on ${\cal N}_u$.

The second step, referred to as {\it injection} requires us to obtain the field
values at the intersection of ${\cal N}_{v+\Delta v}$ with ${\cal N}_u$. This intersection
can be obtained up to second order accuracy by means of equation (\ref{ext_match}),
since
\begin{eqnarray}
v + \Delta v &=& v + \lambda l^0|_{\Gamma} + O(\lambda^2) \, ,  \\
r_i &=& R +  \lambda l^r|_{\Gamma} + O(\lambda^2) \, . 
\end{eqnarray}
From which we can deduce the value of $r_{i}$. Thus,
since the values of the fields are known everywhere on ${\cal N}_u$,
we can interpolate to obtain the values of the outgoing fields at
the intersection point. Finally, by means of the transformation (\ref{ext_match}),
the values of the incoming fields are obtained with which
one can integrate the equations inward.

\end{document}